\documentclass[apj,letterpaper]{emulateapj}
\usepackage{epsfig}

\newcommand{\beq}{\begin{equation}}
\newcommand{\eeq}{\end{equation}}
\newcommand{\bea}{\begin{eqnarray}}
\newcommand{\eea}{\end{eqnarray}}
\newcommand{\trm}[1]{\textrm{#1}}

\def\cd{\trm{ g cm}^{-2}}

\begin{document}

\shortauthors{WEINBERG, BILDSTEN, \& BROWN}
\shorttitle{HYDRODYNAMIC RUNAWAYS IN SUPERBURSTS}

\title{Hydrodynamic Thermonuclear Runaways in Superbursts}
\author{Nevin N.~Weinberg\altaffilmark{1,2}, Lars Bildsten\altaffilmark{2,3}, and Edward F.~Brown\altaffilmark{4}}
\altaffiltext{1}{Astronomy Department and Theoretical Astrophysics Center, 601 Campbell
Hall, University of California, Berkeley, CA 94720; nweinberg@astro.berkeley.edu}
\altaffiltext{2}{Kavli Institute for Theoretical Physics, Kohn Hall, University of California, Santa Barbara, CA 93106; bildsten@kitp.ucsb.edu}
\altaffiltext{3}{Department of Physics, University of California, Santa Barbara, CA 93106}
\altaffiltext{4}{Department of Physics and Astronomy and the Joint Institute of Nuclear Astrophysics, Michigan State University, East Lansing, MI 48824; ebrown@pa.msu.edu} 

\begin{abstract}

We calculate the thermal and dynamical evolution of the surface layers of an accreting neutron star during the rise of a superburst. For the first few hours following unstable $^{12}$C ignition, the nuclear energy release is transported by convection. However, as the base temperature rises, the heating time becomes shorter than the eddy turnover time and convection becomes inefficient. This results in a hydrodynamic nuclear runaway, in which the heating time becomes shorter than the local dynamical time. Such hydrodynamic burning can drive shock waves into the surrounding layers and may be the trigger for the normal X-ray burst found to immediately precede the onset of the superburst in both cases where the {\sl Rossi X-Ray Timing Explorer} was observing. 
\end{abstract}

\keywords{accretion, accretion disks --- nuclear reactions, nucleosynthesis, abundances --- stars: neutron --- X-rays: bursts}

\section{Introduction}
\label{sec:intro} 

Superbursts are hour-long X-ray bursts powered by unstable thermonuclear burning on the surfaces of accreting neutron stars in low mass X-ray binaries. Discovered with long term monitoring campaigns by BeppoSAX and the {\sl Rossi X-Ray Timing Explorer} (RXTE), their recurrence time ($\approx 1 \trm{ yr}$), duration ($\approx \trm{ hours}$), and energies ($\approx 10^{42} \trm{ ergs}$), are $\sim1000$ times greater than that of normal type I X-ray bursts  (see \citealt{Kuulkers:04, Cumming:04a, Strohmayer:06} for reviews). 

The current view is that superbursts are fueled by the unstable burning of a thick layer of $^{12}$C \citep{Cumming:01, Strohmayer:02}, which calculations show is present in the ashes of H/He burning in normal bursts \citep{Brown:98, Schatz:03, Cooper:06}. Fits to superburst lightcurves suggest ignition column depths of $0.5-3\times10^{12} \trm{ g cm}^{-2}$ and energy releases of $2 \times 10^{17} \trm{ ergs g}^{-1}$, implying $^{12}$C mass fractions $X_{12} > 10\%$ \citep{Cumming:06}. For accretion rates $\dot{M} < 0.3 \dot{M}_{\rm Edd}$ (where $\dot{M}_{\rm Edd}$ is the global Eddington accretion rate), a larger $^{12}$C fraction is needed  ($X_{12} \ga 20\%$) if conditions for the thermal instability are reached before the $^{12}$C burns stably away \citep{Cumming:01, Cumming:06}.

Previous theoretical studies have focused on two aspects of the $^{12}$C burning scenario: the thermal structure of the surface layers just before ignition \citep{Cumming:01, Strohmayer:02, Brown:04, Cooper:05, Cumming:06, Cooper:06} and the thermal evolution of the fully burned layers as they cool following the burst \citep{Cumming:01,  Strohmayer:02, Cumming:04b, Cooper:05, Cumming:06}. In this paper, we investigate the rise of the superburst starting from $^{12}$C ignition.  Unlike the cooling studies, which assume the entire layer of fuel burns instantly, we resolve the thermal and dynamical evolution during the burst rise. Understanding the details of the rise is important because it sets the initial conditions for the cooling models. 

The principle result of our study is that burning is so explosive during a superburst rise that the heating time $t_h \equiv (d\ln T_b/ dt)^{-1}$ becomes shorter than the dynamical time $t_d = h / c_s \approx 10^{-6} \trm{ s}$, where $T_b$, $h$, and $c_s$ are the temperature, pressure scale height, and sound speed at the base of the burning layer.  Such a small $t_h$ is possible because the burning takes place in a strongly degenerate layer (cf. \citealt{Woosley:76, Taam:78}). During a normal burst, H/He ignites in a weakly degenerate layer and the burning terminates once radiation pressure dominates at $T_b \approx 2\times10^9 \trm{ K}$; throughout the burn $t_h \ga 10^{-3} \trm{ s}$ (e.g., \citealt{Weinberg:06}).

Once $t_h \la t_d$, convection cannot transport the nuclear energy release and the base undergoes a local thermonuclear runaway, rapidly burning all the $^{12}$C within a thin shell. A hydrodynamic combustion wave forms and propagates into the surrounding fuel, likely driving shock waves into the overlying layers. By contrast, in normal X-ray bursts the energy is transported exclusively by convection and heat diffusion (see \citealt{Bildsten:98}).

Our primary aim here is to describe those ignition conditions that result in a hydrodynamic nuclear runaway. That is, we examine the range of parameters for which the minimum heating time satisfies $t_{h, {\rm min}} < t_d$. We therefore limit our present analysis to the sequence of events leading up to the formation of a combustion wave. In \S~\ref{sec:initial} we describe the initial conditions of our rise models and in \S~\ref{sec:rise} we present our main results. We conclude in \S~\ref{sec:summary} and briefly discuss how the hydrodynamic nature of the burning can influence the observed rise.

\section{Initial Conditions}
\label{sec:initial}

We investigate how the superburst rise depends on three ignition parameters: the column depth $y_b$ at the base of the $^{12}$C layer,  the initial $^{12}$C  mass fraction $X_{12}$, and the composition of the heavy ashes of H/He burning. We assume a neutron star mass and radius of $M=1.4M_\odot$ and $R=10\trm{ km}$ and adopt a plane-parallel approximation with a constant surface gravity $g=(GM/R^2)(1+z) = 2.4\times10^{14}\trm{ cm s}^{-2}$ and redshift $1+z=1.3$. We assume electrons, ions, and photons supply the pressure and calculate the equation of state, volumetric neutrino emissivity $\epsilon_{\nu}$, and thermal conductivity $K$, as in \citet{Brown:04}.

The thermal profile of the accumulated layer is found by integrating the entropy and heat equations
\bea
\label{eq:entropy}
\frac{dF}{dy} &= &\epsilon_{\nu} - \epsilon_{\rm nuc}, \\
\label{eq:heat}
\frac{dT}{dy} &=& \frac{F}{\rho K},
\eea
where $y$ is the rest mass column depth, $F$ the heat flux, $T$ the temperature, $\rho$ the density, and $\epsilon_{\rm nuc}$ the rate of heating from nuclear reactions.  When the accreted layers are in hydrostatic balance, the pressure $p=gy$. We integrate equations (\ref{eq:entropy}) and (\ref{eq:heat}) inwards, setting the outer boundary at $y=10^3\cd$, where we adopt a radiative zero solution.  The flux at the outer boundary has contributions from hot CNO burning $F_{\rm CNO}$, and deep crustal heating $F_{\rm crust}$. We calculate $F_{\rm CNO}$ as in \citet{Cumming:00} and assume the accreted gas has a solar composition. 

For a given choice of  $y_b$, $X_{12}$, and ash composition, we iterate to find the value of $F_{\rm crust}$ that results in unstable $^{12}$C ignition at $y_b$. We define unstable $^{12}$C ignition according to a (numerically solved) one-zone thermal instability criterion $d\epsilon_{\rm nuc} / d\ln T = d\epsilon_{\rm cool} / d\ln T$, where $\epsilon_{\rm cool} = \rho K T/y^2$ (\citealt{Fujimoto:81, Fushiki:87, Cumming:00, Cumming:06}). We change the composition from H/He fuel to ashes of H/He burning at the depth $y_{\rm He}$ where He unstably ignites\footnote{The true value of $y_{\rm He}$ depends on the time since the last normal X-ray burst. This becomes important when modeling the precursor bursts. However, equating $y_{\rm He}$ to the He ignition depth is a reasonable approximation when calculating the $^{12}$C ignition conditions since the accretion timescale at $y_b$ is $\sim 10^3$ times longer than the normal burst recurrence time.} assuming a local accretion rate per unit area $\dot{m}=0.1\dot{m}_{\rm Edd}$, where $\dot{m}_{\rm Edd}$ is the local Eddington accretion rate. The $^{12}$C energy generation rate is given by \citet{Caughlan:88} with screening from \citet{Ogata:93}. This rate only accounts for the energy released in fusing $^{12}$C to $^{24}$Mg and is thus an underestimate if the $^{12}$C burns to iron-group elements or if the heavy (trans-iron) ashes of H/He burning photodisintegrate (\citealt{Schatz:03b}). 
 
\section{The Superburst Rise}
\label{sec:rise}

The rise of a superburst can be divided into three stages: a convective stage, a runaway stage, and a hydrodynamic stage. During the convective stage, convection transports the heat flux from the base $y_b$ to the top of the convective zone $y_c = y_c(T_b)$. We calculate $y_c(T_b)$ in \S~\ref{sec:convective} and show that since the thermal time  $t_{\rm th}(y_c) =C_p y_c^2 / \rho K$ is greater than the heating time $t_h \sim C_p T_b / \epsilon_{\rm nuc}$, the convective zone extends outward to lower pressures as $T_b$ rises ($C_p$ is the specific heat at constant pressure). 

Eventually $T_b$ is so high that $t_h$ is smaller than the eddy turnover time $t_e = h/v_c$, where $v_c =v_c(T_b)$ is the typical velocity of a convective cell. At that point, convection becomes inefficient and the fuel is rapidly consumed within a thin layer near the base in a local thermonuclear runaway. We show in \S~\ref{sec:runaway} that during the runaway, $t_h$ becomes shorter than $t_d$. 
 
\begin{figure}
\begin{center}
\epsfig{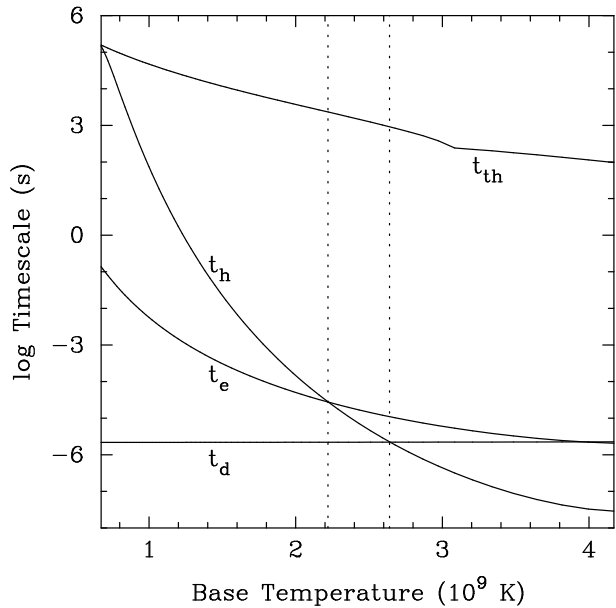}
\end{center}
\caption{Evolution of the heating time $t_h = (d\ln T_b / dt)^{-1}$ during the burst rise as a function of  base temperature $T_b$. Also shown are the thermal time at the top of the convective zone $t_{\rm th}$, the eddy turnover time at the base $t_e$, and the dynamical time at the base $t_d$. The two vertical dotted lines indicate when the burning makes the transition to the runaway stage and the hydrodynamic stage. We assume here an ignition column depth of $10^{12} \cd$ and an initial composition of 20\% $^{12}$C ($X_{12} = 0.2$) and 80\% $^{56}$Fe by mass. \label{fig:timescale}}
\end{figure}

Figure \ref{fig:timescale} shows the evolution of the heating time $t_h = (d\ln T_b/dt)^{-1}$ during a burst rise relative to the thermal time at the top of the convective zone $t_{\rm th}$, the eddy turnover time at the base $t_e$, and the dynamical time at the base $t_d$, for $y_b = 10^{12} \cd$ and $X_{12}=0.2$. Here and below we define the convective, runaway, and hydrodynamic stages according to when $t_e < t_h < t_{\rm th}$,  $t_d < t_h < t_e$,  and $t_h < t_d$, respectively.

\subsection{Convection}
\label{sec:convective} 

At unstable $^{12}$C ignition, $\nu \epsilon_{\rm nuc} \simeq 2 \epsilon_{\rm cool}$, where $\nu = d\ln \epsilon_{\rm nuc} / d\ln T \approx 26$ is the temperature sensitivity of the reaction (see \citealt{Cumming:01}). Thus, comparing the radiative gradient $\nabla_{\rm rad} \equiv (d\ln T/d\ln y)_{\rm rad} \approx \epsilon_{\rm nuc} / \epsilon_{\rm cool}$ to the adiabatic gradient $\nabla_{\rm ad}\equiv (d\ln T/d\ln y)_{\rm ad}\simeq 0.3$, we have at ignition $\nabla_{\rm rad} \simeq 1/13 < \nabla_{\rm ad}$,  and the burning layer is initially stable to convection. During the early stages of burning $t_h \approx t_{\rm th}$ and the heat flux from burning diffuses through the overlying accreted layers. However, the overlying layers only heat up slightly before $\nabla_{\rm rad} = \nabla_{\rm ad}$ and a convective zone forms at the base of the burning layer.

\begin{figure}
\begin{center}
\epsfig{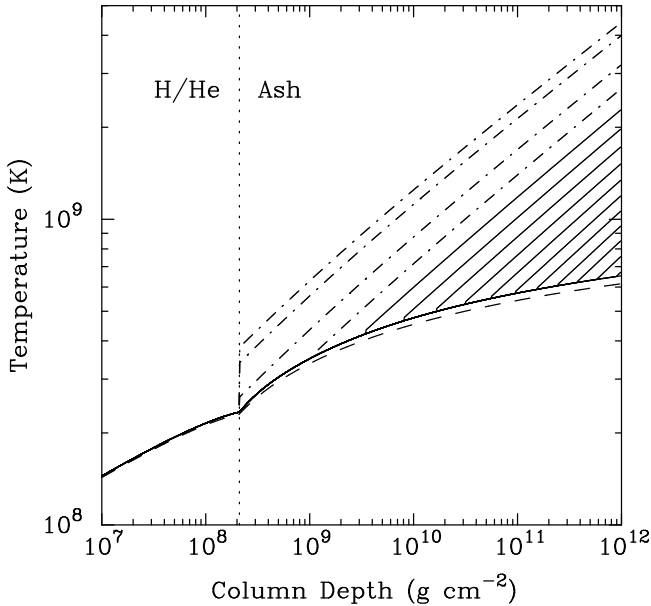}
\end{center}
\caption{Evolution of the thermal profile for an ignition column depth of $10^{12} \cd$ and an initial composition of 20\% $^{12}$C and 80\% $^{56}$Fe by mass. The dashed curve shows the profile at ignition, the solid curve shows the profile at the onset of the convective stage, and the sequence of steep curves are the convective adiabats. The dot-dash curves indicate that the burning at the base is hydrodynamic ($t_h < t_d$). The vertical dotted line shows the location of the H/He--ash interface. \label{fig:Ty}}
\end{figure}

We account for this initial diffusive phase by computing the one-zone rise in temperature $dT_b / dt = \epsilon_{\rm nuc} / C_p$ until $\nabla_{\rm rad} = \nabla_{\rm ad}$. To determine how the overlying layers respond, we integrate equations (\ref{eq:entropy}) and (\ref{eq:heat}) and vary the flux at the top boundary until the solution at the bottom boundary equals $T_b$ at $y_b$. Figure \ref{fig:Ty} shows the change in the thermal profile between ignition (dashed line) and the onset of convection (solid line). Most of the energy release goes into heating up the immediate overlying layers; only a small fraction reaches the H/He layer ($\la 0.1 F_{\rm CNO}$). 

Once formed, the convective zone gradually extends vertically outward to lower pressures. We model the convective evolution using a prescription similar to that of \citet{Weinberg:06}, who modeled convection during normal bursts. We provide a brief overview here, and refer the reader to their paper for further details.

During the convective stage, we assume that the accreted layers are composed of two regions: a completely convective region for $y_c < y < y_b$ (throughout which the ashes of burning are mixed) and a completely diffusive region for $y < y_c$. We define the top of the convective zone $y_c(T_b)$ as the location where the density of the radiative solution just exceeds that of the convective solution. We assume the thermal profile in the convective region follows an adiabat $d\ln T / d\ln y = \nabla_{\rm ad}(y)$. This is reasonable as long as the convective motions are subsonic (making convection efficient), which is true when $t_h > t_e$. The base temperature rises at a rate,
\beq
\label{eq:dTbdt}
\frac{dT_b}{dt} = \frac{\int_{y_c}^{y_b} \epsilon_{\rm nuc} \, dy}{\int_{y_c}^{y_b} C_p (y/y_b)^{\nabla_{\rm ad}} \, dy}.
\eeq
Since the convective zone grows at the heating rate $t_h  < t_{\rm th}(y_c)$, the radiative region cannot thermally adjust to the growing convective region. We therefore assume that heat does not diffuse out through the top of convective zone and the thermal profile in the radiative region is unchanged from its pre-convective state.

The thermal evolution during the convective stage is shown in Figure \ref{fig:Ty}. For $y_b \ga 10^{11} \cd$ and $X_{12} = 0.2$,  $y_c \gg y_{\rm He}$ when $t_h = t_e$ and the burning at the base undergoes a nuclear runaway well before the top of the convective zone ever reaches the H/He layer. It is therefore unlikely that convection will trigger H/He burning before hydrodynamic instabilities develop.

\subsection{Nuclear runaway}
\label{sec:runaway} 

\begin{figure}
\begin{center}
\epsfig{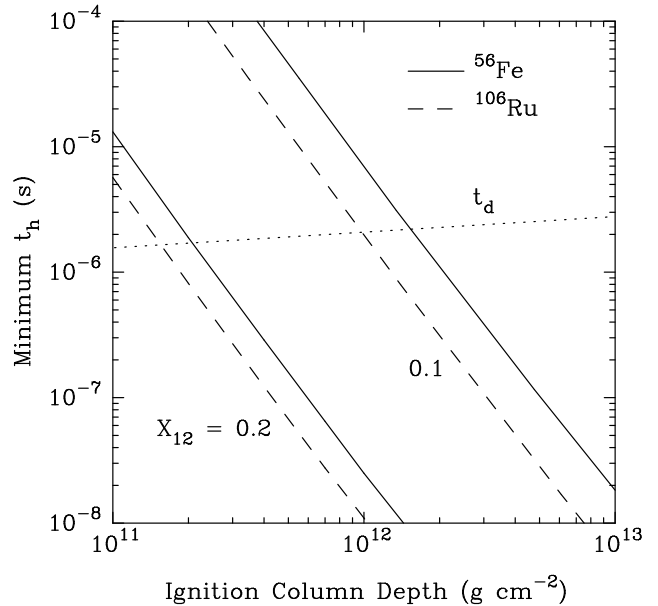}
\end{center}
\caption{Minimum heating time $t_{h, \rm min}$ as a function of the ignition column depth for $X_{12} = 0.1$ and $0.2$. The solid curves are models with iron as the heavy element, the dashed curves for a heavy composition $A=106, Z=44$. Burning becomes hydrodynamic for values of $t_{h, \rm min}$ that lie below the dotted curve, which indicates the dynamical time $t_d$ at the base of the burning layer. \label{fig:tbminy}}
\end{figure}

As Figure \ref{fig:timescale} shows, by the time $T_b \approx 2\times10^9 \trm{ K}$, burning is so rapid that $t_h < t_e$.  Only a small fraction of the $^{12}$C has burned at this point and since the gas is highly degenerate in the burning layer, the density is nearly unchanged from its value at ignition.  No longer able to transport heat via convection, the base undergoes a local thermonuclear runaway.
 
If we continue to naively integrate equation (\ref{eq:dTbdt}) during the runaway, then for ignition column depths greater than a critical value $y_{\rm dyn}$, the heating time $t_h$ becomes even shorter than the dynamical time $t_d$. In reality, a combustion wave must form when $t_h \approx t_d$. In Figure \ref{fig:tbminy} we show the dependence of the minimum heating time $t_{h, {\rm min}}$ on $y_b$, $X_{12}$, and the composition of the burning layer. The nuclear runaway becomes hydrodynamical over the range of superburst ignition parameters. 

We can estimate how $y_{\rm dyn}$ depends on the ignition parameters by assuming that at $t_h = t_{h, {\rm min}}$ half of the $^{12}$C has been burned. The corresponding base temperature $T_{b, 1/2}$ is found by integrating $dX_{12}/dT_b \approx -C_p/E_{\rm nuc}$ up to the midway depletion point, with $C_p$ as in \citet{Cumming:01} and $E_{\rm nuc}$ the nuclear energy release. If we assume $t_h= C_p T_b / \epsilon_{\rm nuc}$ with $\epsilon_{\rm nuc}$ for $^{12}$C burning to $^{24}$Mg as in \citet{Caughlan:88}, and expand the exponential in $\epsilon_{\rm nuc}$ about $T_{b, 1/2}$, we get the minimum heating time in terms of the ignition parameters,
\beq
t_{h, {\rm min}} \approx (3.5\times10^{-8} \trm{ s}) \; 
\left(\frac{2}{g_{14}y_{12}}\right)^{2.9}\left(\frac{2 Y_e}{5 X_{12}}\right)^{9.7}.
\eeq
Here $y_{12} = y_b /10^{12}\cd$, $g_{14} = g/10^{14} \trm{ cm s}^{-2}$ and we assumed the equation of state is that of a relativistic degenerate electron gas with electron fraction $Y_e$. Solving $t_{h, {\rm min}} = t_d$ for $y_{12}$ with $t_d = h / c_s \approx (7\times10^{-6} \trm{ s}) y_{12}^{1/8} Y_e^{1/2}/g_{14}^{7/8}$ then gives
\beq
y_{\rm dyn, 12} = (0.24)
\left(\frac{0.2}{X_{12}}\right)^{3.2}
\left(\frac{Y_e}{0.5}\right)^{3.0}
\left(\frac{2}{g_{14}}\right)^{0.7}.
\eeq
This expression reproduces the more detailed calculation quite well.

The largest thermonuclear energy release possible in a purely convective event $E_{\rm max} = (E_{\rm nuc} X_{12}) (4\pi R^2 y_{\rm dyn})$ is
\beq
E_{\rm max} = (3.4\times10^{41}\trm{ ergs}) 
\left(\frac{0.2}{X_{12}}\right)^{2.2}
\left(\frac{Y_e}{0.5}\right)^{3.0}
\left(\frac{2}{g_{14}}\right)^{0.7}.
\eeq
Events of less energy than this will not lead to hydrodynamic runaways. The observed energy release of superbursts are in the range $0.5-1.4\times10^{42} \trm{ ergs}$ \citep{Kuulkers:04}. 

\section{Summary and Discussion}
\label{sec:summary} 
 
We have shown that the superburst rise evolves through three nuclear burning stages: an hour-long convective stage, a runaway stage, and a hydrodynamic stage. As $T_b$ rises during the convective stage, the heating rate increases and eventually burning is so rapid that $t_h \approx t_e$, resulting in a local nuclear runaway. During the runaway, $t_h$ becomes even shorter than the dynamical time $t_d$, and a hydrodynamic combustion wave forms. 

The combustion wave propagates from the site of the runaway either subsonically as a deflagration or supersonically as a detonation. The issues are similar to those found in theoretical studies of type Ia supernova, in which a C/O white dwarf ignites $^{12}$C at (or near) its center, where the initial density is $\sim 10^9 \trm{ g cm}^{-3}$. In a superburst, the combustion wave is subject to much larger gravitational accelerations. One can show that if the superburst combustion wave is a deflagration, it must always be subject to strong Rayleigh-Taylor instabilities. The correct solution might therefore be a detonation rather than a deflagration. 

A normal X-ray burst is found to immediately precede the onset of the superburst in both cases where the RXTE Proportional Counter Array was observing \citep{Strohmayer:02, Strohmayer:02b}. These precursors may be triggered by hydrodynamic perturbations originating beneath the H/He layer.  In particular, if the combustion wave propagates as a detonation, it will drive shocks into the surrounding layers (see \citealt{Zingale:01}). A deflagration will also generate shocks if the burning front is sufficiently distorted by Rayleigh-Taylor instabilities. The shocks will initially be weak, with overpressures at $y_b$ of $\Delta p/ p \sim E_{\rm nuc} / E_{\rm init} \approx 0.2$, where $E_{\rm init}$ is the initial internal energy of the gas at the ignition depth. However, the shocks steepen as they propagate upwards, becoming quite strong by the time they reach the H/He layer at $y \approx 3\times10^8 \trm{ g cm}^{-2}$. To estimate the strength, approximate the initially weak, upward-propagating, shock as a plane parallel acoustic wave for which $(\Delta p)^2/\rho c_s$ is constant by flux conservation. For $y \ga 10^{10} \trm{ g cm}^{-2}$, relativistic degeneracy pressure dominates and $\rho \propto y^{3/4}$, $c_s \propto y^{1/8}$. Thus, while the shock is weak, the overpressure increases with decreasing column depth as $\Delta p / p \propto y^{-9/16}$. When the shock is strong $\Delta p / p \propto y^{-(1+6\beta)/4}$ where $\beta \simeq 1/(2+[2\gamma/(\gamma-1)]^{1/2})$ (\citealt{Whitham:74}; see also \citealt{Matzner:99}). Here $\gamma =4/3$ which gives $(1+6\beta)/4 = 0.56 \simeq 9/16$ so that over the entire range in $y$ we have 
\beq
\Delta p / p \approx 0.2 (y / y_{12})^{-9/16},
\eeq
and at the H/He layer $\Delta p / p \approx 20 y_{12}^{9/16}$. In an upcoming paper we show that a shock of this strength deposits sufficient entropy into the H/He layer that it can trigger He burning and thereby engender a precursor to the main burst. 

\acknowledgements 

We thank Frank Timmes for helpful conversations. This work was supported by the National Science Foundation under grants AST02-05956,  PHY99-07949, and AST-0507456.

\end{document}